# Efficient resistive memory effect on SrTiO$_3$ by ionic-bombardment

Heiko Gross[1] and Seongshik Oh[2,a)]

[1] Physikalisches Institut, Universität Würzburg, Am Hubland, 97074 Würzburg, Germany

[2] Department of Physics and Astronomy, Rutgers, The State University of New Jersey, 136 Frelinghuysen Road, Piscataway, NJ 08854-8019, USA

(Dated: June 3, 2011)

SrTiO$_3$ is known to exhibit resistive memory effect either with cation-doping or with high-temperature thermal reduction. Here, we add another scheme, ionic-bombardment, to the list of tools to create resistive memory effect on SrTiO$_3$ (STO). In an Ar-bombarded STO crystal, two orders of resistance difference was observed between the high and low resistive states, which is an order of magnitude larger than those achieved by the conventional thermal reduction process. One of the advantages of this new scheme is that it can be easily combined with lithographic processes to create spatially-selective memory effect.

[a)] Electronic mail: ohsean@physics.rutgers.edu

There is growing demand for high-capacity non-volatile memory devices. At present, flash memory, which is based on tunneling in silicon platform, is the most widely used non-volatile memory scheme. However, its limitations in terms of speed and size led researchers to look for alternatives. One of the promising candidates is the resistive memory (RM) effect, which exhibits different resistance values depending on writing voltages[1-3]. Over the past decade, a number of binary and ternary oxides were found to exhibit RM effect[4-9]. Recently, it was also found that RM device is the long sought-after memristor, which existed only in theory for many decades[10-13]. This discovery has made the RM device stand out uniquely among various non-volatile memory candidates.

In this Letter, we show that ionic bombardment can lead to efficient RM effect on pure STO. $SrTiO_3$ is a band insulator but becomes metallic either with cation-doping or with oxygen vacancies[14-17]. This very mechanism that makes insulating STO metallic can also lead to the interesting RM effect with a proper electroforming process. So far two approaches are known to create RM effect on STO. The first is to dope metal ions into STO[9] and the other is to introduce oxygen vacancies into STO by annealing it in vacuum at high temperatures[8, 18]. Another way to introduce oxygen vacancies into STO is the ionic bombardment. Although the ionic bombardment has long been used as a dry-etching technique, called ion-milling, its capability to create controlled oxygen vacancies in STO was recognized only recently[19-22]. Unlike the thermal annealing method, ionic bombardment can create spatially - both vertically and horizontally - confined oxygen vacant layers, and such oxygen vacant layers can lead to interesting insulator-to-metal transition on STO[23]. Now we demonstrate that the same technique can be used to create noticeable RM effect.

In order to create oxygen vacancies, we bombarded Ar ions onto pure STO with beam energy of 66-200 eV and ion current density of 0.5 mA/cm$^2$ in a high vacuum chamber (base pressure, ~5×10$^{-8}$ Torr). It is known that temperature has a significant effect on the formation process of oxygen vacancies; annealing STO in vacuum above ~700 °C creates oxygen vacancies throughout the entire substrate, because the crystal starts to lose oxygen into the vacuum[18]. In our case, the STO temperature never exceeded 300 °C during ion milling, and thus thermal formation of oxygen vacancies is completely suppressed. However, once oxygen vacancies are formed on STO surface by Ar bombardment, they can diffuse deep into the substrate even at this temperature[23]. We monitored the content of oxygen vacancies every half an hour between each Ar bombardment via electric conductance measurement using *in situ* four-point probes[23]. The probes were placed on the sample only during electrical measurement and moved far away during ion milling; the detailed setup can be found elsewhere[24]. For this experiment, we used pure single crystal (001) SrTiO$_3$ substrates (one-side polished, 10×10×0.5 mm$^3$) from MTI and Crystec, *as-received* without any further treatment.

After 20 hours of ion-milling, the entire STO substrate became metallic (sheet resistance of 3.3Ω) with dark blue color[19, 25]. We then exposed the sample to an atmospheric pressure (760 Torr) of pure molecular oxygen for an hour. This led to a thin insulating surface on top of highly conducting STO substrate as shown in Fig. 1. A gold coated tip was placed on the top surface, and the back side of the substrate was grounded. The estimated contact area of the gold tip was around 20 μm$^2$, and the memory effect was measured all in vacuum (<10$^{-7}$ Torr).

In Fig. 2, we demonstrate development of the RM effect on this ion-bombarded sample in four steps. At the beginning, the oxidized STO surface exhibited a diode behavior, characteristic of a Schottky barrier at the gold-STO interface (Fig. 2a). The polarity is consistent with n-type

STO; negative bias on the gold tip repels n-type carriers from the interface, raising the Schottky barrier and this leads to higher resistance, and positive bias attracts n-type carriers toward the interface, lowering the Schottky barrier and the current increases sharply as the bias voltage overcomes the barrier.

However, this diode-like behavior changed irreversibly after large negative voltage was applied. By applying negative voltages up to -14 V, we observed an increase of current that persisted when the bias voltage was reduced to zero as shown in Fig. 2b, implying that the sample became highly conducting after the large negative bias was applied (the compliance current was set to 10 mA). This process can be understood in the following way. Oxygen vacancies are positively charged because oxygen ions are negatively charged with two electrons. A large negative bias on the tip attracts positively charged oxygen vacancies while the Schottky barrier prevents the electrons from shorting the large electric field required to move the oxygen vacancies. But once the oxygen vacancies punch through the thin oxidized surface layer, the Schottky barrier disappears and a highly conducting narrow channel is formed between the tip and the bulk of the metallic STO. This process is called electroforming. The I-V characteristic after the electroforming process shows an ohmic behavior in Fig. 2c (with compliance current of 1mA) at low bias voltage, implying the absence of the Schottky barrier. When the voltage was increased to about +6 V, the current suddenly dropped by two orders of magnitude and the system switched to a high resistance state (HRS). However, unlike the irreversible electroforming process, when a negative voltage of about -6 V was applied, the previous low resistance state (LRS) was resumed, as shown in Fig. 2d. This whole process survived over the entire cycles (>100) we studied.

Although exact mechanism of this reversible RM effect is unknown, it can be explained by opening and closing of nano-scale filamentary channels. During the electroforming process, the large negative bias attracts the positively charged oxygen vacancies to punch through the Schottky barrier forming highly conducting filamentary paths. With positive bias, negatively charged oxygen ions are attracted toward the tip plugging many of the filamentary paths. Some of these highly conducting filamentary paths are open again as negative bias pushes away the oxygen ions. Because not all filamentary paths are closed or opened during each cycle, the switching voltages when these transitions occur can take a range of values as shown in Fig. 3a. These switching voltages tend to decrease as temperature rises (inset of Fig. 3a), indicating the existence of activation barriers against vacancy movement. However, up to the maximum temperature we measured (200 °C), the switching voltages still remained larger than 4 V. This implies that RM devices made out of this scheme will function reliably up to well above room temperature.

Although the observed bipolar switching behaviors are qualitatively similar to those reported previously with other schemes, it is interesting to note that the observed resistance ratio of the high and low resistive states (HRS/LRS) is at least an order of magnitude larger than those reported from the previous oxygen vacant STO samples, which were all prepared by the high temperature annealing process. For a more direct comparison of our own, we prepared a thermally oxygen-deficient STO sample by annealing it up to 770 °C until the sample sheet resistance became similar (~3 Ω) to that of the ion-bombarded sample. The HRS/LRS ratio obtained from this sample (Fig. 3b) was comparable to the values available in the literature for thermally annealed samples[8, 18]. Because of lack of enough statistics, we cannot make a full conclusion out of this comparison, but we speculate that there may be a fundamental origin why the ion-bombarded STO sample exhibited larger RM effect than those prepared by any high temperature annealing processes so far.

Although RM effect can originate from multiple origins[1-3], for oxygen vacant STO samples prepared by either ion-bombardment or thermal annealing there is no doubt that movement of oxygen vacancies is the key player of the RM effect. Therefore, the efficiency of the RM effect for these samples must strongly depend on the mobility of oxygen vacancies. It is known that oxygen vacancies favor to form multivacancy clusters, and depending on the size of the clusters the mobility of the vacancies can significantly vary[26]. However, even if clusters are thermodynamically the preferred phase, single vacancies should still overcome activation barriers and diffuse in order to form clusters with other vacancies. So at finite temperatures, the density of clusters is very likely to be limited by kinetics (diffusion), which gets more effective at high temperatures[23]. Compared with the high-temperature thermal reduction process, which requires sample temperature higher than 700 $^{o}$C, the ion bombardment occurred at much lower temperatures (< 300 $^{o}$C). Therefore, the ion-bombarded sample is expected to have lower density of vacancy clusters than the thermally-annealed samples. With less clusters and more single vacancies, vacancies will have higher mobility, thus better RM effect, in ion-bombarded samples. In addition to the clustering effect, other factors such as thin amorphous surface layer[21-22], surface roughening and higher density of surface vacancies on ion-bombarded STO may have also affected the RM efficiency. However, effects of these other factors are not as clear as that of the vacancy clustering. Further studies will be necessary in order to sort out each of these contributions.

In Fig. 4, we demonstrate how these bipolar switching effects can be used for a memory device by writing and reading HRS and LRS states; positive (negative) 8 V pulse was used to write LRS (HRS) and 1V pulse was used to read the state. Even if this particular demonstration was taken with the thermally reduced sample, whose HRS/LRS ratio was only about ten, the two states were still clearly resolvable over many cycles (>20).

In conclusion, we observed large bipolar switching behaviors on Ar-bombarded $SrTiO_3$ single crystal. The observed resistance ratio between the high and low resistance states was significantly larger than those of thermally-reduced STO samples. Considering that this ionic-bombardment scheme can be easily combined with lithographic techniques such as shadow masking and focused ion beam, it may provide an efficient way to create RM and memristor devices.

**ACKNOWLEDGEMENTS**

We thank Namrata Bansal for assistance during manuscript preparation. This work is supported by IAMDN of Rutgers University, National Science Foundation (NSF DMR-0845464) and Office of Naval Research (ONR N000140910749). H. Gross is also supported by the German academic exchange service (DAAD 221-ISAP D/07/16496).

References

1. A. Kingon, Nature Materials **5**, 251 (2006).

2. A. Sawa, Materials Today **11**, 28 (2008).

3. R. Waser and M. Aono, Nature Materials **6**, 833 (2007).

4. J. R. Jameson, Y. Fukuzumi, Z. Wang, P. Griffin, K. Tsunoda, G. I. Meijer and Y. Nishi, Applied Physics Letters **91**, 112101 (2007).

5. P. Maksymovych, S. Jesse, P. Yu, R. Ramesh, A. P. Baddorf and S. V. Kalinin, Science **324**, 1421 (2009).

6. A. Sawa, T. Fuji, M. Kawasaki and Y. Tokura, Applied Physics Letters **85**, 4073 (2004).

7. D. C. Kim, S. Seo, S. E. Ahn, D. S. Suh, M. J. Lee, B. H. Park, I. K. Yoo, I. G. Baek, H. J. Kim, E. K. Yim, J. E. Lee, S. O. Park, H. S. Kim, U. I. Chung, J. T. Moon and B. I. Ryu, Applied Physics Letters **88**, 202102 (2006).

8. X. B. Yan, Y. D. Xia, H. N. Xu, X. Gao, H. T. Li, R. Li, J. Yin and Z. G. Liu, Applied Physics Letters **97**, 112101 (2010).

9. Y. Watanabe, J. G. Bednorz, A. Bietsch, C. Gerber, D. Widmer, A. Beck and S. J. Wind, Applied Physics Letters **78**, 3738 (2001).

10. D. B. Strukov, G. S. Snider, D. R. Stewart and R. S. Williams, Nature **453**, 80 (2008).

11. J. Borghetti, G. S. Snider, P. J. Kuekes, J. J. Yang, D. R. Stewart and R. S. Williams, Nature **464**, 873 (2010).

12. J. J. Yang, F. Miao, M. D. Pickett, D. A. A. Ohlberg, D. R. Stewart, C. N. Lau and R. S. Williams, Nanotechnology **20**, 215201 (2009).

13. D.-H. Kwon, K. M. Kim, J. H. Jang, J. M. Jeon, M. H. Lee, G. H. Kim, X.-S. Li, G.-S. Park, B. Lee, S. Han, M. Kim and C. S. Hwang, Nature Nanotechnology **5**, 148 (2010).

14. O. N. Tufte and P. W. Chapman, Physical Review **155**, 796 (1967).

15. K. v. Benthem, C. Elsasser and R. H. French, Journal of Applied Physics **90**, 6156 (2001).

16. R. Astala and P. D. Bristowe, Computational Materials Science **22**, 81 (2001).

17. W. Luo, W. Duan, S. G. Louie and M. L. Cohen, Physical Review B **70**, 214109 (2004).

18. K. Szot, W. Speier, G. Bihlmayer and R. Waser, Nature Materials **5**, 312 (2006).

19. D. Kan, O. Sakata, S. Kimura, M. Takano and Y. Shimakawa, Japanese Journal of Applied Physics **46**, L471 (2007).

20. H. Y. Hwang, A. Ohtomo, N. Nakagawa, D. A. Muller and J. L. Grazul, Physica E: Low-dimensional Systems and Nanostructures **22**, 712 (2004).

21. D. W. Reagor and V. Y. Butko, Nature Materials **4**, 593 (2005).

22. D. Kan, T. Terashima, R. Kanda, A. Masuno, K. Tanaka, S. Chu, H. Kan, A. Ishizumi, K. U, Y. Shimakawa and M. Takano, Nature Materials **4**, 816 (2005).

23. H. Gross, N. Bansal, Y.-S. Kim and S. Oh, arXiv:1104.5522.

24. H. Gross, M.S. Thesis, Rutgers, The State University of New Jersey, 2009.

25. H. D. Hwang, Nature Materials **4**, 803 (2005).

26. D. D. Cuong, B. Lee, K. M. Choi, H.-S. Ahn, S. Han and J. Lee, Physical Review Letters **98**, 115503 (2007).

**FIGURE CAPTIONS**

**Fig. 1** (Color online) Measurement setup. Oxygen ions and vacancies are represented by O and V, respectively. Exposure of the ion-bombarded STO, which is highly conductive, to ambient oxygen forms a thin insulating surface layer. By applying a voltage to the contact, the resistance between the contact and the back side of the sample can be changed.

**Fig. 2** (Color online) Creating bipolar switching behaviors on ion-bombarded STO in four steps. (a) I-V characteristic of gold-tip/oxidized-surface-layer/conductive-STO shows a Schottky diode behavior. (b) Electroforming step with compliance current of 10 mA: applying a large negative bias makes the positive vacancies accumulate below the gold contact, and then the system switches to the low resistive state. (c) Ohmic characteristic near zero bias in the low resistive state with compliance current of 1 mA. (d) Full hysteresis loop: sufficient high positive (negative) bias voltage switches the system to the high (low) resistive state. Note the two orders of resistance difference between the two states.

**Fig. 3** (Color online) (a) Switching loops for the ion-bombarded STO: through the entire measured cycles, resistance ratios between HRS and LRS states stayed close to a hundred. Inset: switching voltage as a function of sample temperature: the error bars represent the range of switching voltages. (b) Switching loop for the high-temperature annealed sample. HRS/LRS resistance ratio is approximately ten, about ten times smaller than the ion-bombarded sample.

**Fig. 4** (Color online) Writing and reading of the LRS/HRS states (thermally annealed sample). Positive (negative) 8 V pulse was used to write LRS (HRS) and 1V pulse was used to read the resistance state.

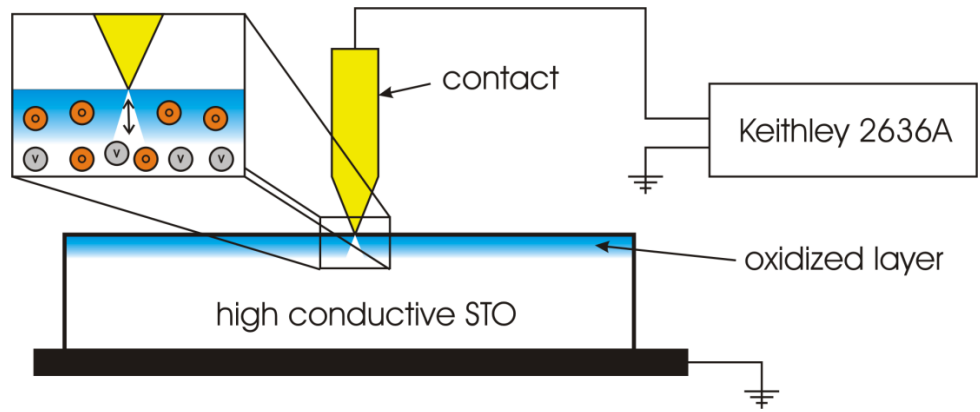

**Fig. 1**

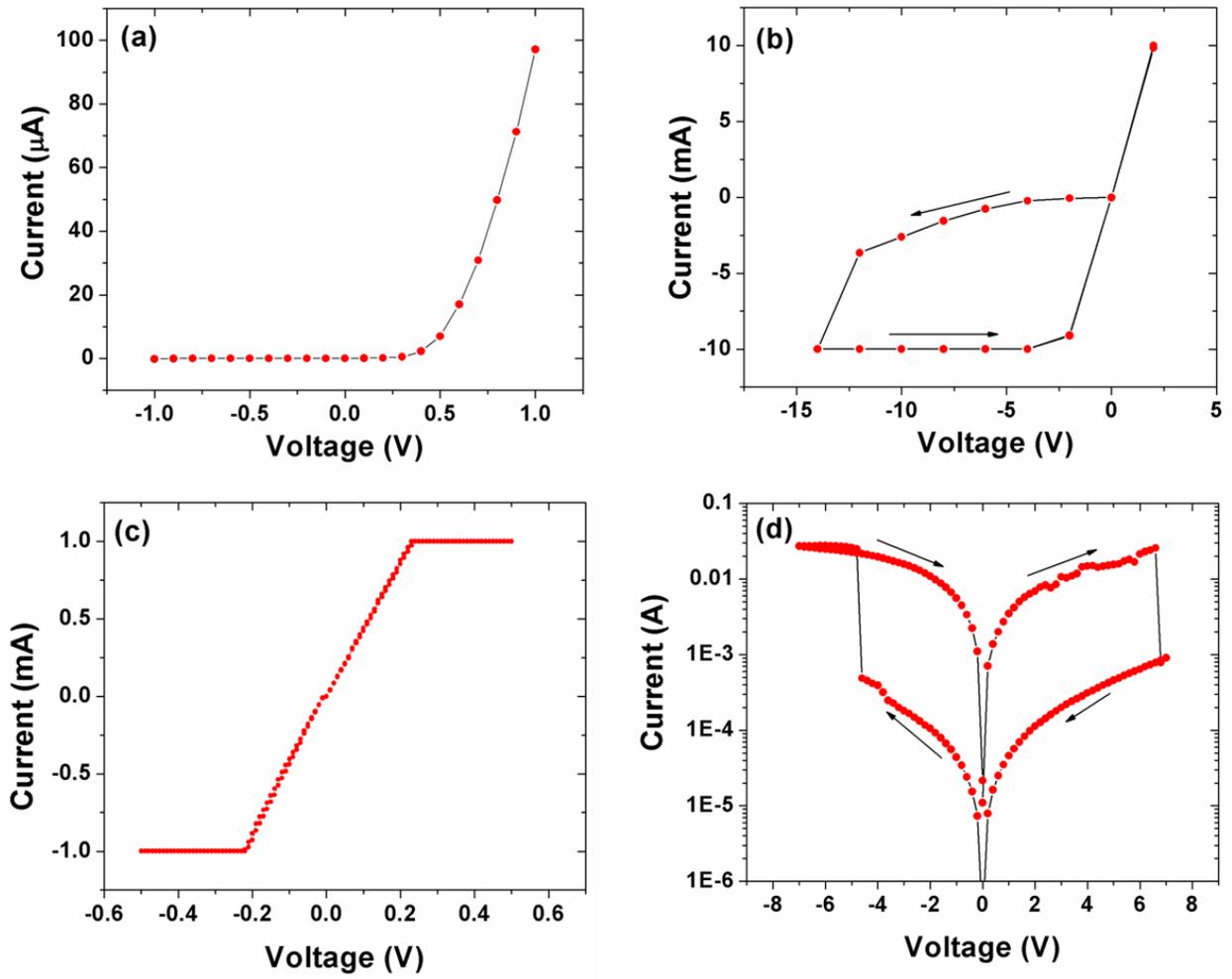

**Fig. 2**

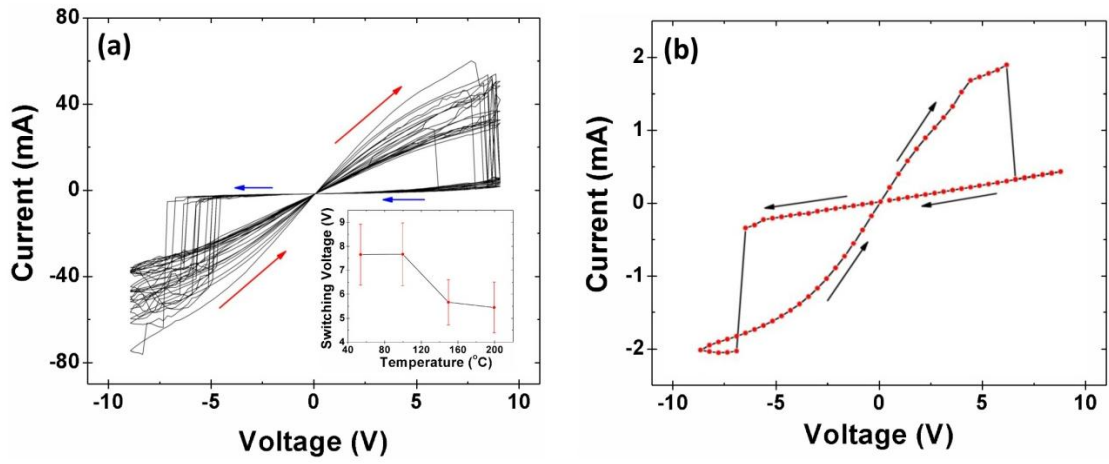

**Fig. 3**

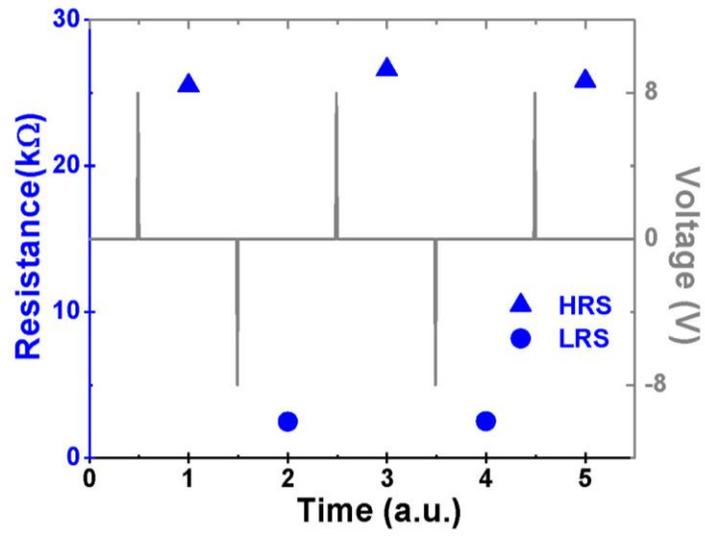

**Fig. 4**